\documentclass[12pt]{article}

\usepackage{graphicx}
\begin{document}

\begin{center}
{\bf Nonlinearly charged AdS black holes, extended phase space thermodynamics and Joule–Thomson expansion } \\
\vspace{5mm} S. I. Kruglov
\footnote{E-mail: serguei.krouglov@utoronto.ca}
\underline{}
\vspace{3mm}

\textit{Department of Physics, University of Toronto, \\60 St. Georges St.,
Toronto, ON M5S 1A7, Canada\\
Department of Chemical and Physical Sciences, University of Toronto,\\
3359 Mississauga Road North, Mississauga, Ontario L5L 1C6, Canada} \\
\vspace{5mm}
\end{center}
\begin{abstract}
We study thermodynamics and phase transitions of magnetically charged by nonlinear electrodynamics (NED) Anti-de Sitter (AdS) black holes in an extended phase space. The cosmological constant is treated as a thermodynamic pressure whereas the black hole mass is
considered such as the chemical enthalpy. The horizon thermodynamics of black holes shows an analogy with the Van der Walls liquid–gas system.
We define a quantities which are conjugated to the parameter of NED and a magnetic charge. It is shown that the first law of thermodynamics and the generalized Smarr relation hold. The critical exponents are the same as in the Van der Waals system. The Joule--Thomson adiabatic expansion of NED-AdS black holes is investigated because of thermodynamic behavior of black holes. The Joule--Thomson coefficient, the inversion and isenthalpic curves are discussed. The minimum of inversion temperature and the corresponding event horizon radius are obtained.
\end{abstract}

\section{Introduction}

It was proven that a black hole is a thermodynamics system \cite{Bardeen}, \cite{Jacobson}, \cite{Padmanabhan}  where the area of the black hole is treated such as the entropy and the surface gravity is its temperature \cite{Bekenstein}, \cite{Hawking}. This can provide
the insight on  black hole gravity and to establish the theory of quantum gravity. The studding AdS spacetime with a negative cosmological constant shown the possibility of phase transitions in black holes \cite{Page}. In addition, considering gravity with a negative cosmological constant allowed to formulate a holographic picture with black holes being a system that is dual to conformal field theories \cite{Maldacena}, \cite{Witten}, \cite{Witten1}. The holography can solve some problems in quantum chromodynamics \cite{Kovtun} and physics of condensed matter  \cite{Kovtun1}, \cite{Hartnoll}. Recently, it was suggested that the cosmological constant may be considered as a thermodynamics pressure in the black hole that is a conjugate to a volume. It was demonstrated that the phase transitions in black holes are similar to liquid-gas thermodynamics \cite{Dolan}, \cite{Kubiznak}, \cite{Mann}, \cite{Teo}.
Here, we investigate nonlinear electrodynamics (NED), proposed in \cite{Kr0}, \cite{Kr2}, coupled to gravity with AdS spacetime in extended phase space thermodynamics. Models of NED were introduced to remove singularities and to take into account quantum gravity corrections to classical Maxwell fields. The valuable NED at weak-field limit gives the Maxwell fields. The first example of valuable NED is Born--Infeld electrodynamics
\cite{Born}. The model of NED considered has an attractive features which are similar to Born--Infeld electrodynamics. The reason to consider NED \cite{Kr0} is in its simplicity. The mass and metric functions are expressed in the form of simple elementary functions while in in Born--Infeld model these functions of the hypergeometric type. Therefore, the study of this NED-AdS black hole allows us to make reliable calculations.
Born--Infeld  electrodynamics coupled to gravity with the negative cosmological constant was studied in \cite{Fernando}-\cite{Miskovic} where it was observed an analogy with Van der Waals fluids. We study the Joule--Thomson adiabatic thermal expansion of NED-AdS black holes with heating and cooling regimes. There are many papers devoted to Joule--Thomson expansion of AdS black holes \cite{Aydiner}, \cite{Yaraie}, \cite{Mo}, \cite{Rizwan}, \cite{Chabab}, \cite{Mirza}, \cite{Bi} (and others). In the pioneer work \cite{Aydiner} authors studied Joule–-Thomson effects for charged (by Maxwell electrodynamics) AdS black holes. Corrections of the quintessence on the Joule--Thomson expansion of the Reissner--Nordstr\"{o}m AdS black hole were investigated in \cite{Yaraie}. The authors of \cite{Mo} studied effects of the dimensionality on the Joule--Thomson expansion by considering the case of d-dimensional linearly charged AdS black holes. Authors of \cite{Rizwan} investigated the Joule--Thomson effects for AdS black holes with a global monopole. Joule--Thomson effects for charged AdS black holes in $f(R)$ gravity were studied in \cite{Chabab}.  In \cite{Mirza}, the thermal stability and Joule--Thomson expansion of the quasitopological black holes were studied. The Joule--Thomson expansion of Born--Infeld AdS black holes was investigated in the extended phase space \cite{Bi}.

In the present paper we study the Joule–-Thomson expansion of AdS black holes in the framework of NED.
The Joule--Thomson expansion takes place when the enthalpy, which is a black hole mass, is constant. When the pressure decreases for the expanding black holes then at the inversion pressure $P_i$, a cooling and heating transition occurs \cite{Aydiner}. Here, we obtain NED-AdS black hole solution and study thermodynamics, $P-V$ criticality and Joule--Thomson expansion.

In section 2 we find the NED-AdS metric function for magnetized black holes and corrections to the Reissner{Nordstr\"{o}m solution. The asymptotic as $r\rightarrow 0$ was obtained showing the de Sitter core. In section 3 we study first law of black hole thermodynamics in the extended phase space where a negative cosmological constant is a pressure. We obtain the thermodynamic magnetic potential and the thermodynamic conjugate to the NED coupling. It is demonstrated that the generalized Smarr relation holds. In section 4 the critical specific volume, critical temperature and critical pressure are obtained. The expression for Gibbs free energy is found and analysed as well as critical exponents. The Joule--Thomson adiabatic expansion of NED-AdS black holes is studied in Section 5. The Joule--Thomson coefficient, the inversion and isenthalpic curves are discussed. The minimum of inversion temperature and the corresponding event horizon radius are obtained. Section 6 is devoiced to conclusion and summary. In Appendix the Kretschmann scalar is calculated and depicted in Fig. 11.

The units with $c=1$, $\hbar=1$, $k_B=1$ are explored.

\section{NED-AdS black hole solution}

The action of NED-AdS theory in general relativity is defined as
\begin{equation}
I=\int d^{4}x\sqrt{-g}\left(\frac{R-2\Lambda}{16\pi G_N}+\mathcal{L}(\mathcal{F}) \right),
\label{1}
\end{equation}
where $G_N$ is the Newton constant, the negative cosmological constant being $\Lambda=-3/l^2$ with AdS radius $l$. We use the NED Lagrangian proposed in \cite{Kr0} (see also \cite{Kr2})
\begin{equation}
{\cal L}(\mathcal{F}) =-\frac{{\cal F}}{1+\sqrt{2|\beta{\cal F}|}}.
\label{2}
\end{equation}
The Lorentz invariant is ${\cal F}=F^{\mu\nu}F_{\mu\nu}/4=(B^2-E^2)/2$, where $E$ and $B$ are the electric and magnetic induction fields, respectively. At $\beta=0$ in Eqs. (1) and (2), we arrive at action of charged by linear Maxwell electrodynamics AdS black holes.
From action (1) we obtain the gravitation and electromagnetic field equations
\begin{equation}
R_{\mu\nu}-\frac{1}{2}g_{\mu \nu}R+\Lambda g_{\mu \nu} =8\pi G_N T_{\mu \nu},
\label{3}
 \end{equation}
\begin{equation}
\partial _{\mu }\left( \sqrt{-g}\mathcal{L}_{\mathcal{F}}F^{\mu \nu}\right)=0,
\label{4}
\end{equation}
where$\mathcal{L}_{\mathcal{F}}=\partial \mathcal{L}( \mathcal{F})/\partial \mathcal{F}$.
The stress tensor of electromagnetic fields is given by
\begin{equation}
 T_{\mu\nu }=F_{\mu\rho }F_{\nu }^{~\rho }\mathcal{L}_{\mathcal{F}}+g_{\mu \nu }\mathcal{L}\left( \mathcal{F}\right).
\label{5}
\end{equation}
Let us consider spacetime with the spherical symmetry,
\begin{equation}
ds^{2}=-f(r)dt^{2}+\frac{1}{f(r)}dr^{2}+r^{2}\left( d\theta
^{2}+\sin ^{2}\theta d\phi ^{2}\right).
\label{6}
\end{equation}
Then the tensor $F_{\mu\nu}$ possesses the radial electric field $F_{01}=-F_{10}$ and radial
magnetic field $F_{23}=-F_{32}=q_m\sin(\theta)$ with the magnetic charge $q_m$. The stress tensor is diagonal, $T_{0}^{~0}=T_{r}^{~r}$ and $T_{\theta}^{~\theta}=T_{\phi}^{~\phi}$. The metric function in Eq. (6) is \cite{Bronnikov}
\begin{equation}
f(r)=1-\frac{2m(r)G_N}{r},
\label{7}
\end{equation}
and the mass function is given by
\begin{equation}
m(r)=m_0+\int_{0}^{r}\rho (r)r^{2}dr.
\label{8}
\end{equation}
Here, we included in Eq. (8) the Schwarzschild mass $m_0$ which is the integration constant and $\rho(r)$ being the energy density. It is worth mentioning that models with the electrically charged black holes having Maxwell's weak-field limit possess singularities \cite{Bronnikov}. Therefore, we consider magnetic black holes with $\mathcal{F}=q_m^2/(2r^4)$.

From Eq. (5) (electric charge $q_e=0$) the energy density, including the term which is due to the negative cosmological constant, becomes
\begin{equation}
\rho=\frac{q_m^2}{2r^2(r^2+q_m\sqrt{\beta})}-\frac{3}{2G_Nl^2}.
\label{9}
\end{equation}
 Making use of Eqs. (8) and (9) one finds the mass function
\begin{equation}
m(r)=m_0+\frac{q_m^{3/2}}{2\beta^{1/4}}\arctan\left(\frac{r}{\sqrt{q_m}\beta^{1/4}}\right)-\frac{r^3}{2G_Nl^2}.
\label{10}
\end{equation}
We define the black hole magnetic mass
\begin{equation}
m_M=\int_0^\infty \frac{q_m^2}{2(r^2+ q_m\sqrt{\beta})}dr=\frac{\pi q_m^{3/2}}{4\beta^{1/4}}.
\label{11}
\end{equation}
According to Eq. (11) the magnetic energy becomes infinite at the Maxwell limit $\beta=0$. This reflects the fact that the
total magnetic energy of magnetic monopole (as well as electric energy of point-like charges) is infinite. In our NED, coupling $\beta$
smoothes singularities. The same situation takes place in Born--Infeld electrodynamics. With the help of Eqs. (7) and (10) we obtain the metric function
\begin{equation}
f(r)=1-\frac{2m_0G_N}{r}-\frac{q_m^{3/2}G_N}{\beta^{1/4}r}\arctan\left(\frac{r}{\sqrt{q_m}\beta^{1/4}}\right)+\frac{r^2}{l^2}.
\label{12}
\end{equation}
At large value of coupling, $\beta\rightarrow \infty$, the metric function (12) corresponds to Schwarzschild-AdS black hole case.
From Eq. (12), without the cosmological constant ($l\rightarrow \infty$), we find the metric function as $r\rightarrow \infty$
\begin{equation}
f(r)=1-\frac{2(m_0+m_M)G_N}{r}+\frac{q_m^2G_N}{r^2}-\frac{q_m^3\sqrt{\beta}G_N}{3r^4}+\mathcal{O}(r^{-6})~~~\mbox{as}~r\rightarrow \infty.
\label{13}
\end{equation}
Equation (13) shows corrections to the Reissner--Nordstr\"{o}m solution. According to Eq. (13) we can make the identification of
the ADM mass of the black hole with the total mass $M\equiv m_0+m_M$ which is the sum of the Schwarzschild mass $m_0$ and the magnetic mass $m_M$. Making use of Eq. (12), at $m_0=0$, one obtains the asymptotic as $r\rightarrow 0$
\begin{equation}
f(r)=1-\frac{q_mG_N}{\sqrt{\beta}}+\frac{G_Nr^2}{3\beta}+\frac{r^2}{l^2}+\mathcal{O}(r^{4})~~~\mbox{as}~r\rightarrow 0,
\label{14}
\end{equation}
which has a de Sitter core. In accordance with Eq. (14) the metric function is regular (at $m_0=0$) because $f(r)$ is finite
at the origin $r=0$. But the Kretschmann scalar as $r\rightarrow 0$ approaches to infinity (see Appendix). As a result, there is a space-time singularity at $r=0$. According to Fig. 11 the Kretschmann scalar goes to infinity rapidly at smaller NED parameter $\beta$.
The plot of metric function (12) is represented in Fig. 1 at $m_0=0$, $G_N=1$, $l=10$.
\begin{figure}[h]
\includegraphics  {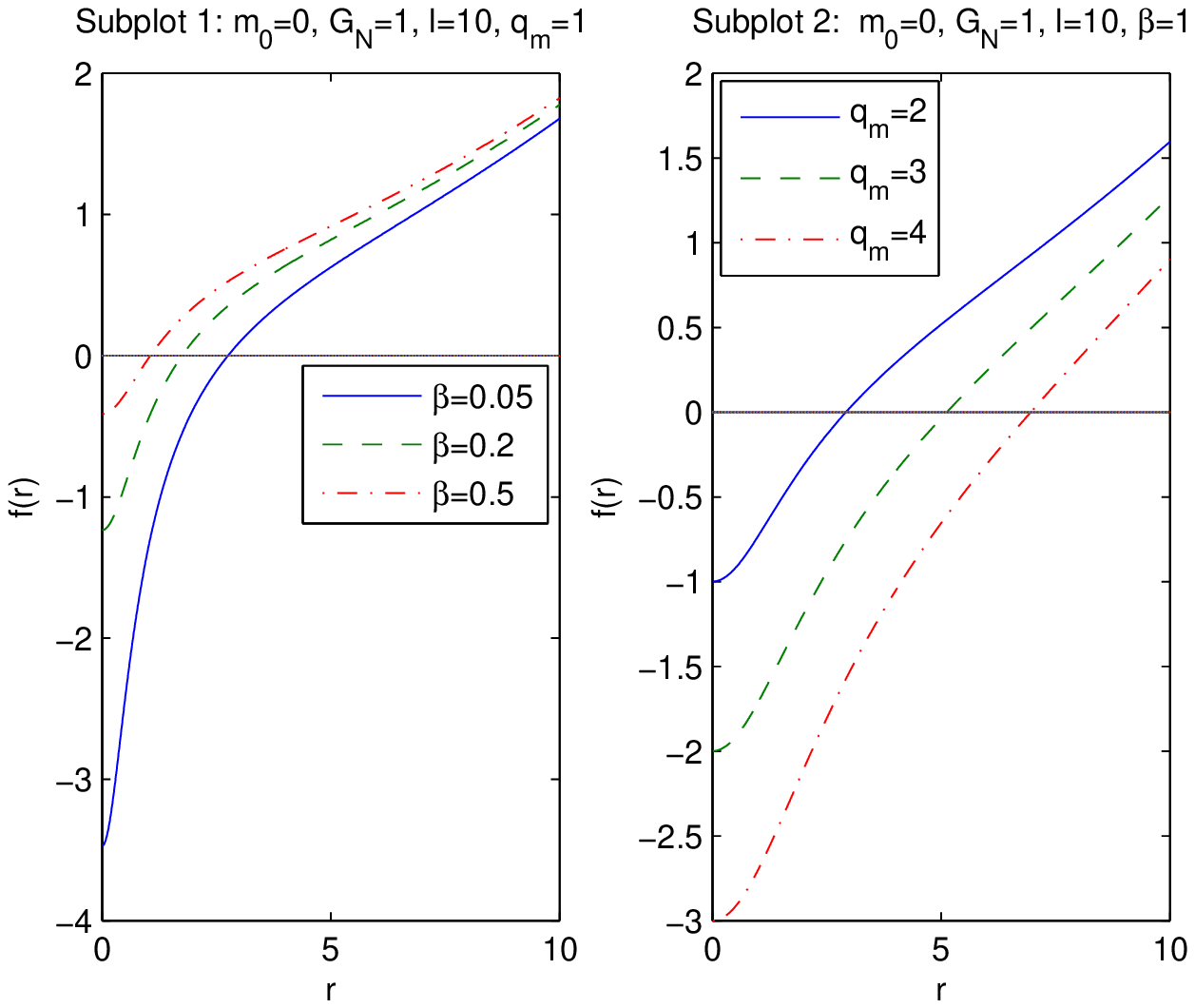}
\caption{\label{fig.1} The plots of the function $f(r)$ at $m_0=0$, $G_N=1$, $l=10$ shows the existence of only one horizon. The metric function is finite at the origin $r=0$. According to subplot 1, if NED constant $\beta$ increases then the event horizon radius decreases. Subplot 2 shows that when magnetic charge $q_m$ increases the event horizon radius also increases.}
\end{figure}
In accordance with Fig. 1 black holes possess only one horizon ($r_+$). When coupling $\beta$ increases at constant $q_m$, the event horizon radius decreases. But if magnetic charge $q_m$ increases at constant $\beta$, the event horizon radius increases.

\section{The Smarr relation and first law of black hole thermodynamics}

To formulate the generalized first law of black hole thermodynamics one has to introduce the pressure which is associated with a
negative cosmological constant $\Lambda$ \cite{Kastor}, \cite{Dolan1}, \cite{Cvetic}, \cite{Kubiznak1}. It is worth noting that there is a problematic issue of the interpretation of bulk pressure \cite{Kubiznak2}.
The first law of black hole thermodynamics is given by $dM=T\delta S+\Omega dJ+\Phi dq$, where the black hole mass is $M$, charge is $q$, and angular momentum is $J$. But here the pressure-volume term $PdV$ is absent. To include the term $PdV$ in the first law of black hole thermodynamics we need the pressure which was associated with a negative cosmological constant $\Lambda$ giving a positive vacuum pressure.  As a result, the generalized first law of black hole thermodynamics is $dM=TdS+VdP+\Omega dJ+\Phi dq$ \cite{Kastor}, \cite{Dolan1}, \cite{Cvetic}. The generalized first law of black hole mechanics becomes the first law of ordinary thermodynamics if we interpret $M$ as a chemical enthalpy \cite{Kastor}, $M=U+PV$, with $U$ being the internal energy.

One can obtain the Smarr formula from the first law of black hole thermodynamics by using the Euler scaling argument \cite{Smarr} (see also \cite{Kastor}). Exploring the dimensional analysis, with $G_N=1$, we find $[M]=L$, $[S]=L^2$, $[P]=L^{-2}$, $[J]=L^2$, $[q_m]=L$, $[\beta]=L^2$. Treating $\beta$ as a thermodynamic variable and taking into account the Euler’s theorem \cite{Mann}, we obtain the mass
\begin{equation}
M=2S\frac{\partial M}{\partial S}-2P\frac{\partial M}{\partial P}+2J\frac{\partial M}{\partial J}+q_m\frac{\partial M}{\partial q_m}+2\beta\frac{\partial M}{\partial \beta}.
\label{15}
\end{equation}
The thermodynamic conjugate to NED parameter $\beta$ is $\partial M/\partial \beta\equiv {\cal B}$ and the black hole entropy $S$, volume $V$ and pressure $P$ are  \cite{Myers}, \cite{Myers1}
\begin{equation}
S=\pi r_+^2,~~~V=\frac{4}{3}\pi r_+^3,~~~P=-\frac{\Lambda}{8\pi}=\frac{3}{8\pi l^2}.
\label{16}
\end{equation}
 Making use of Eq. (12) and equation $f(r_+)=0$, with $r_+$ being the event horizon radius, at $G_N=1$, one finds the black hole mass
\begin{equation}
M=\frac{r_+}{2}+\frac{r_+^3}{2l^2}+\frac{\pi q_m^{3/2}}{4\beta^{1/4}}-\frac{q_m^{3/2}}{2\beta^{1/4}}\arctan\left(\frac{r_+}{\sqrt{q_m}\beta^{1/4}}\right).
\label{17}
\end{equation}
When $\beta\rightarrow \infty$ the mass function (17) becomes Schwarzschild-AdS black hole mass, and
at the limit $\beta\rightarrow 0$ we obtain the mass function (for a linear case) of Maxwell-AdS magnetic black hole
\[
M_l=\frac{r_+}{2}+\frac{r_+^3}{2l^2}+\frac{ q_m^2}{2r_+}.
\]
From Eq. (17), for non-rotating black hole ($J=0$), we obtain
\[
dM=\left(\frac{1}{2}+\frac{3r_+^2}{2l^2}-\frac{q_m^2}{2(r^2+q_m\sqrt{\beta})}\right)dr_+
-\frac{r_+^3}{l^3}dl
\]
\[
+\left(\frac{3\pi \sqrt{q_m}}{8\beta^{1/4}}-\frac{3\sqrt{q_m}}{4\beta^{1/4}}\arctan\left(\frac{r_+}{\sqrt{q_m}\beta^{1/4}}\right)+
\frac{q_mr_+}{4(r_+^2+q_m\sqrt{\beta})}\right)dq_m
\]
\begin{equation}
+\left(-\frac{\pi q_m^{3/2}}{16\beta^{5/4}}+\frac{q_m^{3/2}}{8\beta^{5/4}}\arctan\left(\frac{r_+}{\sqrt{q_m}\beta^{1/4}}\right)+
\frac{q_m^2r_+}{8\beta(r_+^2+q_m\sqrt{\beta})}\right)d\beta.
\label{18}
\end{equation}
The Hawking temperature is defined by
\begin{equation}
T=\frac{f'(r)|_{r=r_+}}{4\pi},
\label{19}
\end{equation}
where $f'(r)=\partial f(r)/\partial r$.
Making use of  Eqs. (12), (19) and equation $f(r_+)=0$ at $G_N=1$, one finds the Hawking temperature
\begin{equation}
T=\frac{1}{4\pi}\biggl(\frac{1}{r_+}+\frac{3r_+}{l^2}-\frac{q_m^2}{r_+(r_+^2+ q_m\sqrt{\beta})}\biggr).
\label{20}
\end{equation}
At $\beta=0$ in (20) we have the Hawking temperature of Maxwell-AdS black hole while at $\beta\rightarrow\infty$, the Hawking temperature of
Schwarzschild-AdS black hole. By virtue of Eqs. (16), (18) and (20) we obtain the first law of black hole thermodynamics
\begin{equation}
dM = TdS + VdP + \Phi_m dq_m + {\cal B}d\beta.
\label{21}
\end{equation}
The thermodynamic magnetic potential $\Phi_m$ and the thermodynamic conjugate to the NED parameter $\beta$ are defined by
\[
\Phi_m =\frac{3\pi \sqrt{q_m}}{8\beta^{1/4}}-\frac{3\sqrt{q_m}}{4\beta^{1/4}}\arctan\left(\frac{r_+}{\sqrt{q_m}\beta^{1/4}}\right)+
\frac{q_mr_+}{4(r^2+q_m\sqrt{\beta})},
\]
\begin{equation}
{\cal B}=-\frac{\pi q_m^{3/2}}{16\beta^{5/4}}+\frac{q_m^{3/2}}{8\beta^{5/4}}\arctan\left(\frac{r_+}{\sqrt{q_m}\beta^{1/4}}\right)+
\frac{q_m^2r_+}{8\beta(r_+^2+q_m\sqrt{\beta})}.
\label{22}
\end{equation}
To obtain the Smarr relation we need the quantity (vacuum polarization) ${\cal B}$. The plot of potential $\Phi_m$ versus $r_+$ is shown in Fig. 2.
\begin{figure}[h]
\includegraphics {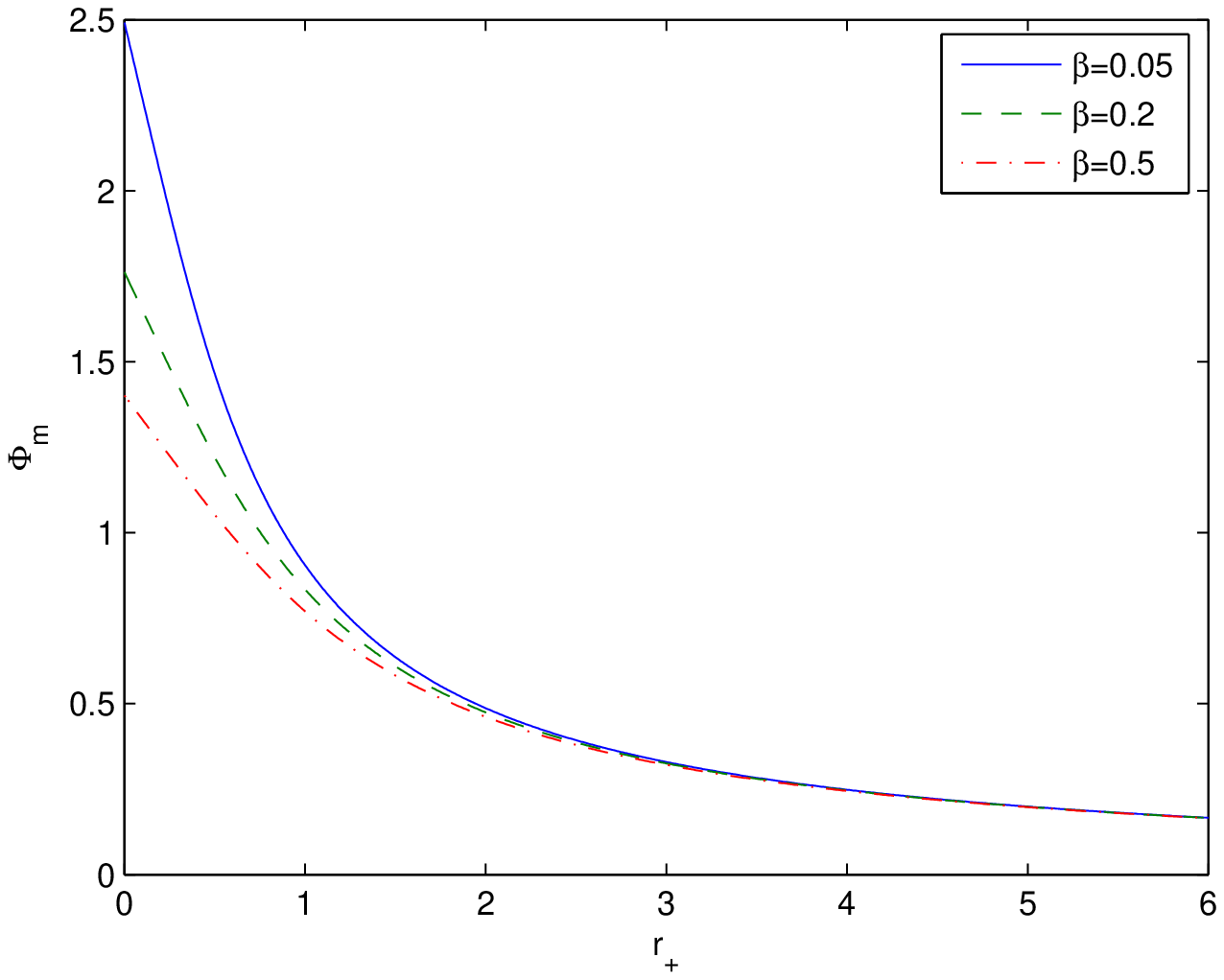}
\caption{\label{fig.2} The plots of the function $\Phi_m$ vs. $r_+$ at $q_m=1$. The solid curve is for $\beta=0.05$, the dashed curve corresponds to $\beta=0.2$, and the dashed-dotted curve corresponds to $\beta=0.5$. The magnetic potential is finite at $r_+=0$ and vanishes at $r_+\rightarrow \infty$. When coupling $\beta$ increases the magnetic potential $\Phi_m$ decreases.}
\end{figure}
In accordance with Fig. 2 if the NED parameter $\beta$ increases the magnetic potential decreases. At $r_+\rightarrow \infty$ the magnetic potential $\Phi_m$ vanishes ($\Phi_m(\infty)=0$) and at $r_+ = 0$ it is finite.
\begin{figure}[h]
\includegraphics{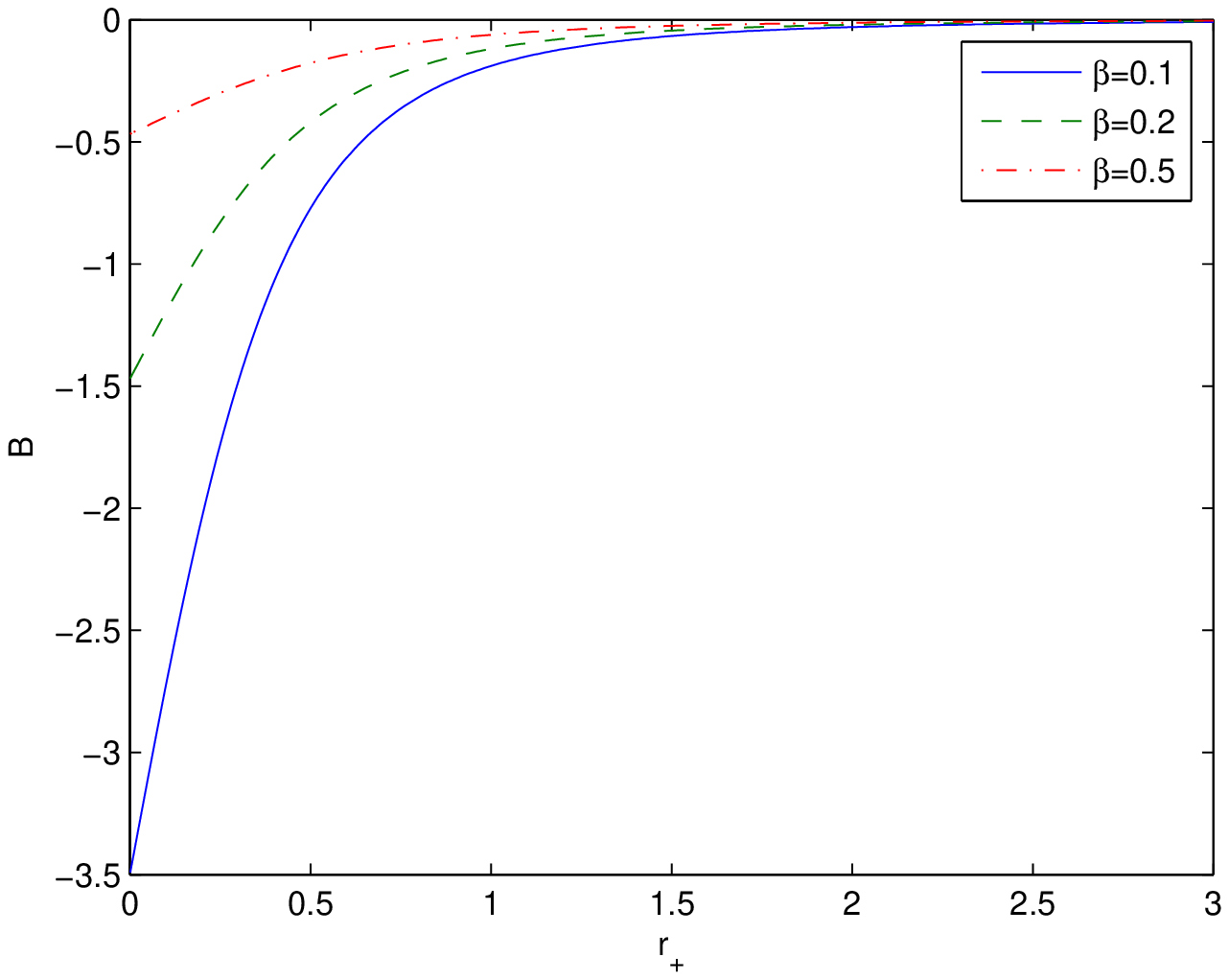}
\caption{\label{fig.3} The plots of vacuum polarization ${\cal B}$ vs. $r_+$ at $q_m=1$. The solid curve is for $\beta=0.1$, the dashed curve corresponds to $\beta=0.2$, and the dashed-dotted curve corresponds to $\beta=0.5$. The function ${\cal B}$ becomes zero at $r_+\rightarrow \infty$ and finite at $r_+=0$.  When NED parameter $\beta$ increases the $|{\cal B}|$ decreases.}
\end{figure}
The plot of ${\cal B}$ versus $r_+$ is depicted in Fig. 3. When $r_+ = 0$ the vacuum polarization ${\cal B}$ becomes finite. According to Fig. 3 if coupling $\beta$ increases $|{\cal B}|$ decreases and as $r_+\rightarrow \infty$ it vanishes (${\cal B}(\infty)=0$).

Taking into account Eqs. (16), (17) and (22), one can check the validity of the generalized Smarr formula
\begin{equation}
M=2ST-2PV+q_m\Phi_m+2\beta{\cal B}.
\label{23}
\end{equation}
Thermodynamics of Born--Infeld-AdS model in extended phase space was investigated in \cite{Mann1}, \cite{Zou}, \cite{Hendi}, \cite{Hendi1}, \cite{Zeng}.

\section{Black hole thermodynamics}

By virtue of Eq. (20) we find the equation of state of the black hole
\begin{equation}
P=\frac{T}{2r_+}-\frac{1}{8\pi r_+^2}+\frac{q_m^2}{8\pi r_+^2(r_+^2+q_m\sqrt{\beta})}.
\label{24}
\end{equation}
At the limit $\beta\rightarrow \infty$ in Eq. (24), we obtain Schwarzschild-AdS black hole equation of state.
If we put $\beta=0$ in Eq. (24), it becomes the equation of state for a charged  Maxwell-AdS black hole \cite{Kubiznak1}.
Comparing the equation of state of charged AdS black hole with the Van der Waals equation, one has to identify the specific volume $v$
with $2l_Pr_+$ \cite{Kubiznak1}, where $l_P=\sqrt{G_N}=1$. Equation (24) becomes
\begin{equation}
P=\frac{T}{v}-\frac{1}{2\pi v^2}+\frac{2q_m^2}{\pi v^2(v^2+4 q_m\sqrt{\beta})},
\label{25}
\end{equation}
which mimics the behavior of the Van der Waals fluid. One can obtain critical points (the inflection point in the $P-v$ diagram) by equations
\[
\frac{\partial P}{\partial v}=-\frac{T}{v^2}+\frac{1}{\pi v^3}-\frac{8q_m^2(v^2+2q_m\sqrt{\beta})}{\pi v^3(v^2+4 q_m\sqrt{\beta})^2}=0,
\]
\begin{equation}
\frac{\partial^2 P}{\partial v^2}=\frac{2T}{v^3}-\frac{3}{\pi v^4}-\frac{8q_m^2(24\beta q_m^2+18v_c^2q_m\sqrt{\beta}+5v^4)}{\pi v^4(v^2+4 q_m\sqrt{\beta})^3}=0.
\label{26}
\end{equation}
Making use of Eq. (26) one finds the equation for the critical points
\begin{equation}
8q_m^2\left(3v_c^4+6q_m\sqrt{\beta}v_c^2+8\beta q_m^2\right)-\left(v_c^2+4 q_m\sqrt{\beta}\right)^3=0.
\label{27}
\end{equation}
From Eq. (26) we obtain the equations for the critical temperature and pressure
\begin{equation}
T_c=\frac{1}{\pi v_c}-\frac{8q_m^2\left(v_c^2+2q_m\sqrt{\beta}\right)}{\pi v_c\left(v_c^2+4q_m \sqrt{\beta}\right)^2},
\label{28}
\end{equation}
\begin{equation}
P_c=\frac{1}{2\pi v_c^2}-\frac{2q_m^2\left(3v_c^2+4q_m\sqrt{\beta}\right)}{\pi v_c^2\left(v_c^2+4q_m\sqrt{\beta}\right)^2}.
\label{29}
\end{equation}
Approximate solutions to Eq. (27) and critical temperatures and pressures are given in Table 1.
\begin{table}[ht]
\caption{Critical values of the specific volume and temperature at $q_m=1$}
\centering
\begin{tabular}{c c c c c c c c }\\[1ex]
\hline
$\beta$ & 0.1 & 0.3 & 0.5 & 0.7 & 0.9 & 1.1 & 1.3  \\[0.5ex]
\hline
$v_{c}$ &4.55 & 4.26 & 4.02 & 3.80 & 3.58 & 3.34 & 3.06 \\[0.5ex]
\hline
$T_{c}$ &0.045 & 0.047 & 0.048 & 0.050 & 0.051 & 0.052 & 0.054 \\[0.5ex]
\hline
$P_{c}$ &0.0036 & 0.0040 & 0.0042 & 0.0045 & 0.0048 & 0.0051 & 0.0055 \\[0.5ex]
\hline
\end{tabular}
\end{table}
The plots of $P-v$ diagrams are depicted in Figs. 4 and 5.
\begin{figure}[h]
\includegraphics {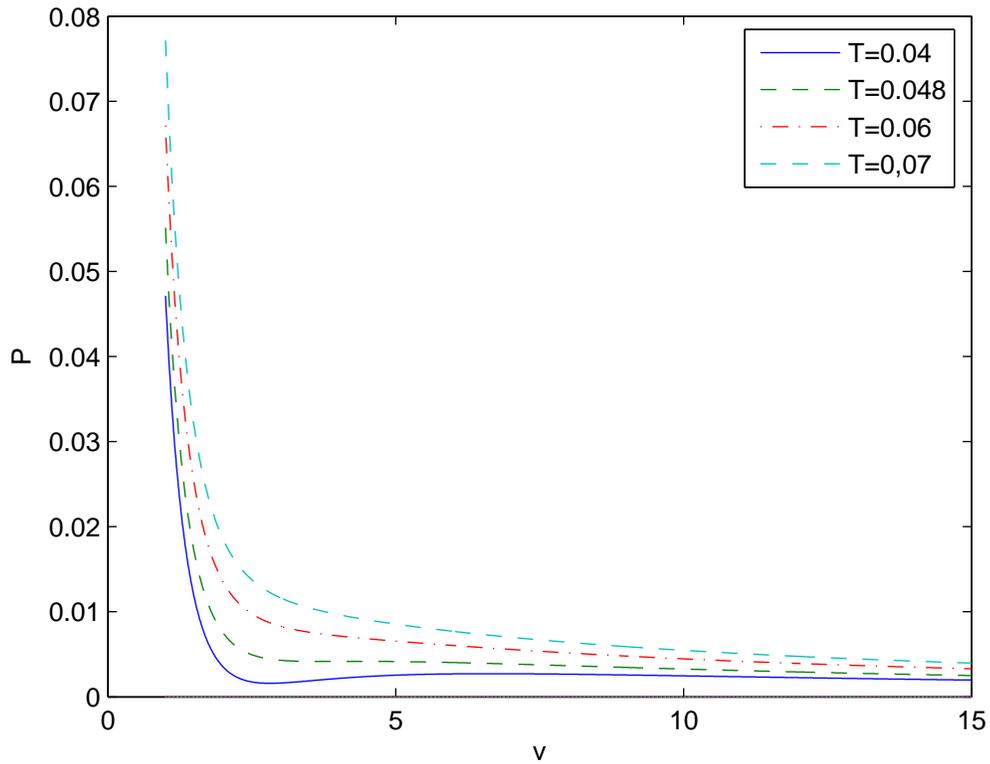}
\caption{\label{fig.4} The plots of the function $P$ vs. $v$ at $q_m=1$, $\beta=0.5$. This equation of state mimics the Van der Waals
fluid behavior. The critical isotherm corresponds to $T_{c}\approx 0.048$ with the inflection point.}
\end{figure}
\begin{figure}[h]
\includegraphics {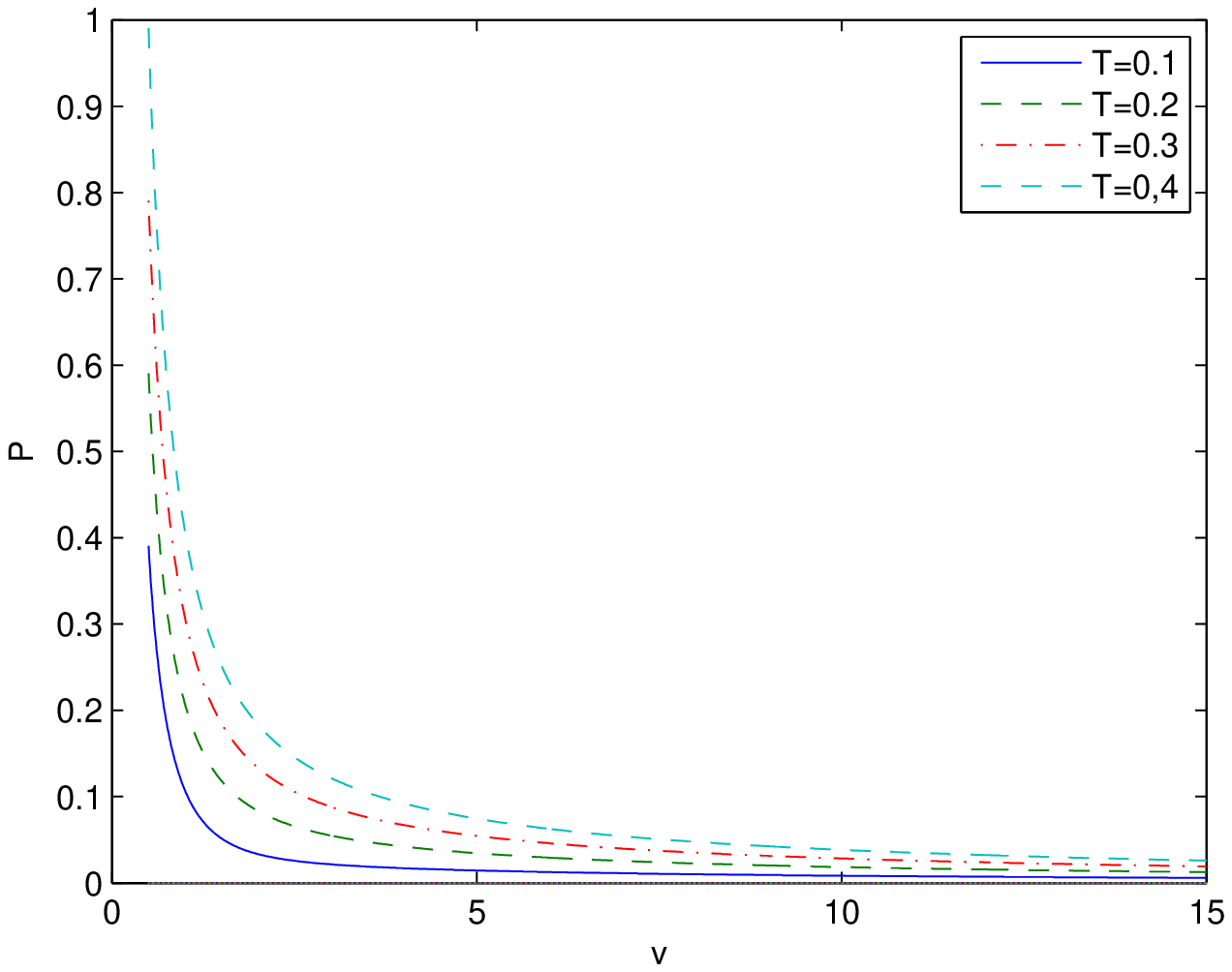}
\caption{\label{fig.5} The plots of the function $P$ vs. $v$ at $q_m=1$, $\beta=0.5$ for $T=0.1$, $0.2$, $0.3$ and $0.4$. These diagrams show non-critical behavior of equation of state. The critical point is ($T_{c}\approx 0.048$, $P_c\approx 0.00424$). There are not the inflection points in diagrams.}
\end{figure}
In accordance with Fig. 4 at $q_m=1$, $\beta=0.5$ the critical value for specific volume is $v_{c}\approx 4.02$ ($T_{c}=0.048$).  Non-critical behaviour of $P-v$ diagrams for $T=0.1, 0.2, 0.3$ and $0.4$ is presented in Fig. 5.
The critical ratio follows from Eqs. (28) and (29),
\begin{equation}
\rho_c=\frac{P_cv_c}{T_c}=\frac{(v_c^2+4q_m\sqrt{\beta})^2-4q_m^2(3v_c^2+4q_m\sqrt{\beta})}{2[(v_c^2+4q_m\sqrt{\beta})^2
-8q_m^2(v_c^2+2q_m\sqrt{\beta})]}.
\label{30}
\end{equation}
Making use of Eq. (30), at $\beta=0$ ($v_c^2=24q_m^2$), one obtains the ratio $\rho_c=3/8$ which is valid for a Van der Waals fluid.
From Eq. (30) we find at strong coupling regime that the critical ratio approaches to $1/2$, $\lim_{\beta\rightarrow\infty} \rho_c =1/2$.
At large values of $\beta$, there are not real solutions to Eq. (27). This is because we come to the case of Schwarzschild-AdS black holes with Hawking--Page phase transitions between thermal radiation and large black holes.


The Gibbs free energy with a fixed charge, NED parameter $\beta$ and pressure, where $M$ is treated as a chemical enthalpy, is given by
\begin{equation}
G=M-TS.
\label{31}
\end{equation}
Making use of Eqs. (17), (20) and (31), at $G_N=1$, we find
\begin{equation}
G=\frac{r_+}{4}-\frac{2\pi r_+^3P}{3}+\frac{\pi q_m^{3/2}}{4\beta^{1/4}}-\frac{q_m^{3/2}}{2\beta^{1/4}}\arctan\left(\frac{r_+}{\sqrt{q_m}\beta^{1/4}}\right)+\frac{q_m^2 r_+}{4(r_+^2+q_m\sqrt{\beta})}.
\label{32}
\end{equation}
At $\beta\rightarrow0$ Eq. (32) becomes the Gibbs free energy of Maxwell-AdS black hole and at $\beta\rightarrow\infty$, it corresponds to Gibbs free energy of Schwarzschild-AdS black hole. The plots of $G$ versus $T$, where according to Eq. (24) $r_+$ is a function of $P$ and $T$, is depicted in Fig. 6.
\begin{figure}[h]
\includegraphics {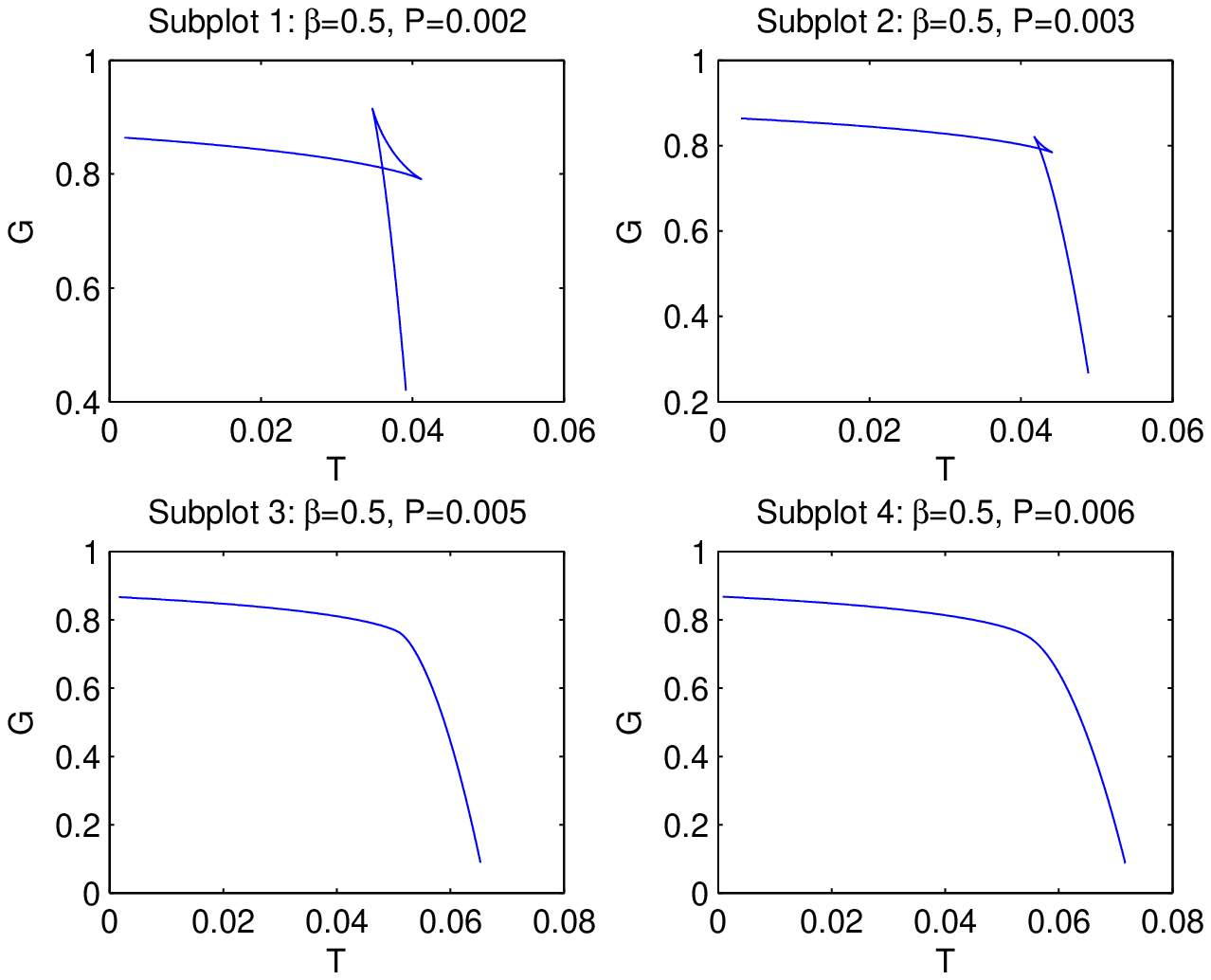}
\caption{\label{fig.6} The plots of the Gibbs free energy $G$ vs. $T$ with $q_m=1$, $\beta=0.5$. Subplots 1 and 2 show the critical 'swallowtail' behavior with first-order phase transitions between small and large black holes. Subplots 3 and 4 correspond to the case $P>P_c$ ($P_c\approx 0.00424$) with non- critical behavior of the Gibbs free energy.}
\end{figure}
We analyzed, as an example, the behavior of $G$ with $\beta=0.5$ and $v_c\approx 4.0206$, $T_c\approx 0.0483$, $P_c\approx 0.00424$.
In subplots 1 and 2 there are critical points with first-order phase transitions between small and large black holes. The Gibbs free energy at $P<P_c$ possesses 'swallowtail' behavior. In subplots 3 and 4 in Fig. 6 for $P>P_c$ the behavior of the Gibbs free energy is similar to the Hawking--Page behavior for Schwarzschild-AdS case.

\subsection{Critical exponents}

In the vicinity of critical points, critical exponents characterize the behaviour of physical quantities, that do not depend on details of the system. Making use of Eqs. (27), (28) and (29) we obtain critical values of the specific volume, temperature and pressure for small $\beta$
\begin{equation}
v_c=2\sqrt{6}q_m+{\cal O}(\beta),~~~T_c=\frac{1}{3\sqrt{6}\pi q_m}+{\cal O}(\beta),~~~P_c=\frac{1}{96\pi q_m^2}+{\cal O}(\beta).
\label{33}
\end{equation}
The critical points (33) at $\beta=0$ are similar to charged AdS black hole points \cite{Mann1}. The critical ratio becomes
\begin{equation}
\rho_c=\frac{3}{8}+{\cal O}(\beta),
\label{34}
\end{equation}
where the value $\rho_c=3/8$ holds for the Van der Waals fluid.
The behavior of specific heat at the constant volume is characterized by parameter $\alpha$
\begin{equation}
C_v=T\frac{\partial S}{\partial T}\propto |t|^{-\alpha},
\label{35}
\end{equation}
where $t=(T-T_c)/T_c$.
The expressions for difference of the large and small black hole volume on the given isotherm $v_l-v_s$, isothermal compressibility $\kappa_T$, $|P-P_c|$ on the critical isotherm $T = T_c$ are given by
\begin{equation}
\eta=v_l-v_s\propto |t|^\beta,~~\kappa_T=-\frac{1}{v}\frac{\partial v}{\partial P}|_T\propto |t|^{-\gamma},~~|P-P_c|\propto |v-v_c|^\delta.
\label{36}
\end{equation}
One can obtain, following the avenue of \cite{Mann1}, parameters as follows
\begin{equation}
\alpha=0,~~~~\beta=\frac{1}{2},~~~~\gamma=3,~~~~\delta=3,
\label{37}
\end{equation}
which are identical with values of critical exponents in the Born--Infeld-AdS theory.

\section{Joule--Thomson expansion of NED-AdS black holes}

The Joule--Thomson expansion is isenthalpic so that the black hole mass $M$, which is the enthalpy, is constant during the expansion. The Joule--Thomson thermodynamic coefficient is introduced to study cooling-heating phases under the adiabatic expansion
\begin{equation}
\mu_J=\left(\frac{\partial T}{\partial P}\right)_M=\frac{1}{C_P}\left[ T\left(\frac{\partial V}{\partial T}\right)_P-V\right]=\frac{(\partial T/\partial r_+)_M}{(\partial P/\partial r_+)_M}.
\label{38}
\end{equation}
In fact, the Joule--Thomson coefficient $\mu_J$ is the slope in $P-T$ diagrams.
The sign of $\mu_J$ is changed at the inversion temperature $T_i$ when $\mu_J(T_i)=0$. If the initial temperature during the expansion is higher than inversion temperature $T_i$, the final temperature decreases corresponding to the cooling phase ($\mu_J>0$). But when the initial temperature is lower than $T_i$, the final temperature increases for this heating phase ($\mu_J<0$). From Eq. (38) and equation $\mu_J(T_i)=0$ we obtain the inversion temperature
\begin{equation}
T_i=V\left(\frac{\partial T}{\partial V}\right)_P=\frac{r_+}{3}\left(\frac{\partial T}{\partial r_+}\right)_P.
\label{39}
\end{equation}
The inversion temperature can be considered as a borderline between cooling and heating process. The inversion temperature
line crosses points in maxima of $P-T$ diagrams where their slope is changed and these points separate cooling and heating phases
of black holes \cite{Yaraie}, \cite{Mo}, \cite{Rizwan}. Equation of black hole equation state (24) can be rewritten as
\begin{equation}
T=\frac{1}{4\pi r_+}+2P r_+-\frac{q_m^2}{4\pi r_+(r_+^2+q_m\sqrt{\beta})}.
\label{40}
\end{equation}
At $\beta=0$ Eq. (40) becomes the equation of state for Maxwell-AdS balack holes.
Making use of Eq. (17) and the equation for pressure $P=3/(8\pi l^2)$ one obtains
\begin{equation}
P=\frac{3}{4\pi r_+^3}\left[M-\frac{r_+}{2}-\frac{\pi q_m^{3/2}}{4\beta^{1/4}}+\frac{q_m^{3/2}}{2\beta^{1/4}}\arctan\left(\frac{r_+}{\sqrt{q_m}\beta^{1/4}}\right)\right].
\label{41}
\end{equation}
Equations (40) and (41) define the $P-T$ isenthalpic diagrams in the parametric form depicted in Fig. 7. The inversion $P_i-T_i$ curve goes through maxima of isenthalpic curves.
 \begin{figure}[h]
\includegraphics {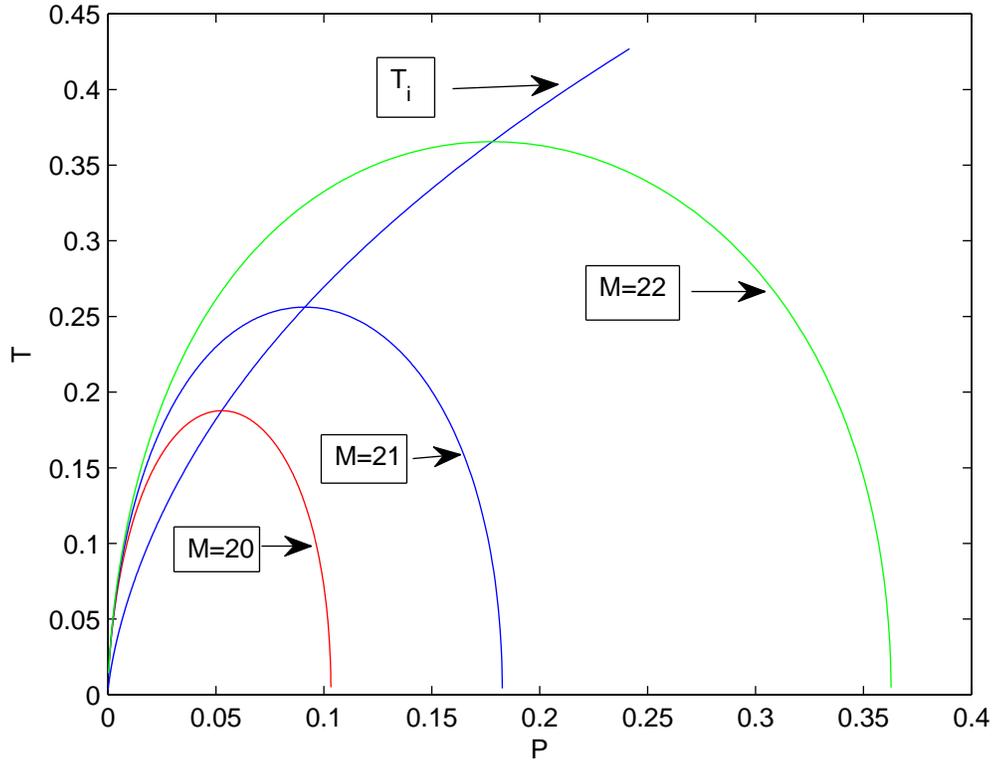}
\caption{\label{fig.7} The plots of the temperature $T$ vs. pressure $P$  and the inversion temperature $T_i$ with $q_m=10$, $\beta=1$.
The inversion $P_i-T_i$ curve goes through maxima of isenthalpic curves. The isenthalpic $P-T$ diagrams correspond to masses $M=20,21,22$ from bottom to top. When black hole masses increase the inversion temperature $T_i$ increases.}
\end{figure}
From Eqs. (39) and (40) we obtain the equation for inversion pressure $P_i$ versus event horizon radius
\begin{equation}
P_i=\frac{q_m^2\left(3r_+^2+2q_m\sqrt{\beta}\right)}{8\pi r_+^2\left(r_+^2+q_m\sqrt{\beta}\right)^2}-\frac{1}{4\pi r_+^2}.
\label{42}
\end{equation}
Making use of Eqs. (40) and (42) we obtain the inversion temperature
\begin{equation}
T_i=\frac{q_m^2\left(2r_+^2+q_m\sqrt{\beta}\right)}{4\pi r_+\left(r_+^2+q_m\sqrt{\beta}\right)^2}-\frac{1}{4\pi r_+}.
\label{43}
\end{equation}
Setting $P_i=0$ in Eq. (42), one finds the corresponding event horizon radius $r_{min}$ and minimum inversion temperature
\begin{equation}
r_+^{min}=\frac{1}{2}\sqrt{q_m\left(3q_m-4\sqrt{\beta}+\sqrt{q_m(9q_m-8\sqrt{\beta})}\right)},
\label{44}
\end{equation}
\begin{equation}
T_i^{min}=\frac{q_m^2r_{min}}{8\pi (r_{min}^2+q_m\sqrt{\beta})^2}.
\label{45}
\end{equation}
The plot of the minimum inversion temperature $T_i^{min}$ versus coupling $\beta$ is depicted in Fig. 8.
\begin{figure}[h]
\includegraphics {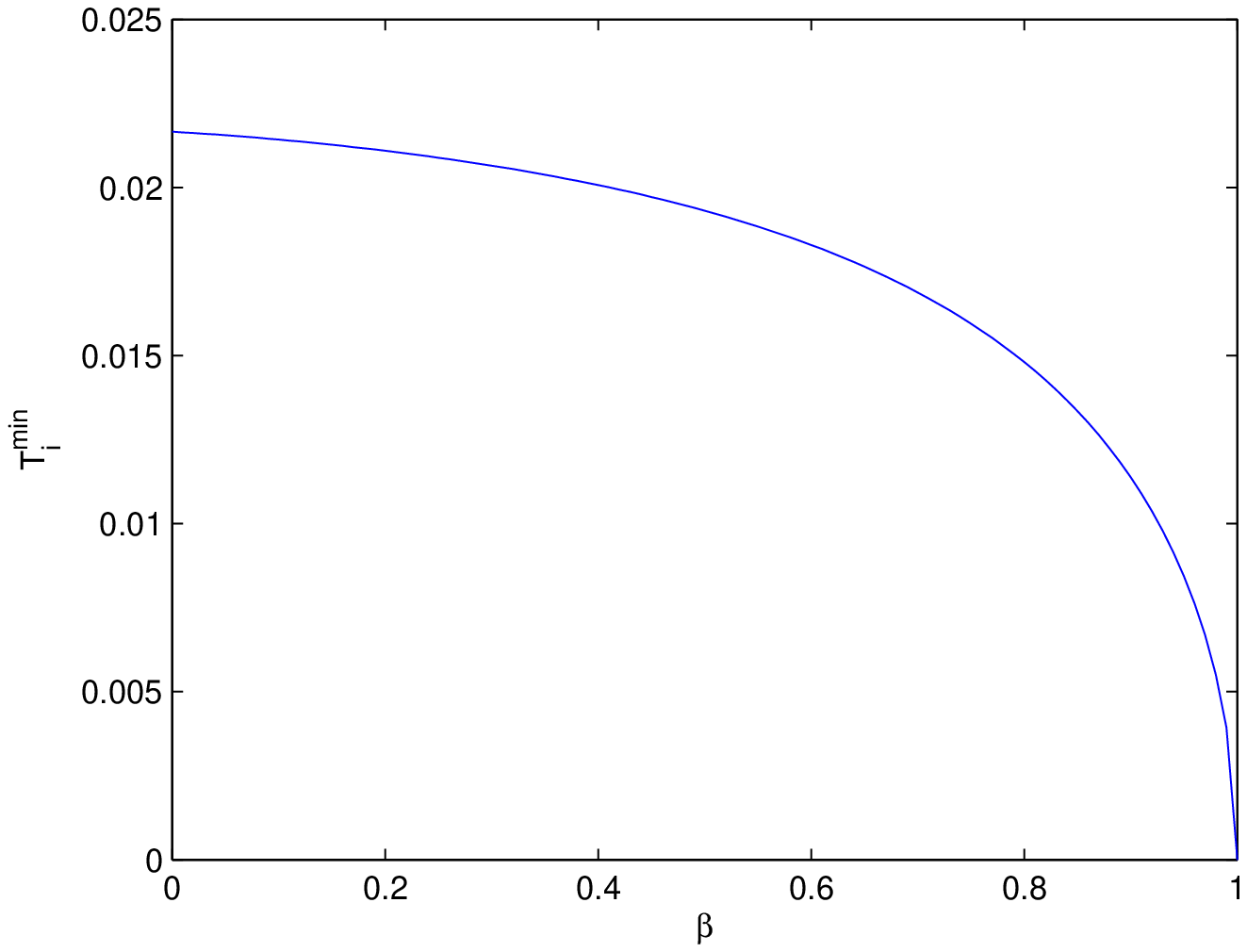}
\caption{\label{fig.8} The plot of the minimum inversion temperature $T_i^{min}$ vs. $\beta$ for $q_m=1$. The minimum inversion temperature decreases with increasing the coupling $\beta$. At $\beta=1$ the minimum inversion temperature $T_i^{min}$ vanishes.}
\end{figure}
Figure 8 shows that the minimum inversion temperature decreases with $\beta$. At the particular values $q_m=\beta=1$ the minimum inversion temperature becomes zero.
Making use of Eqs. (44) and (45) at $\beta=0$ we obtain the minimum of inversion temperature for Maxwell-AdS magnetic black holes
\begin{equation}
T_i^{min}=\frac{1}{6\sqrt{6}\pi q_m}, ~~~~r_+^{min}=\frac{\sqrt{6}q_m}{2}.
\label{46}
\end{equation}
From Eq. (33) for $\beta=0$ and Eq. (46), we obtain the relation $T_i^{min}=T_c/2$. The same equality holds for electrically charged AdS black holes \cite{Aydiner}. In our case $\beta\neq 0$ we have $T_i^{min}\neq T_c/2$.
Equations (42) and (43) are parametric equations for inversion temperature $T_i$ versus $P_i$ which is depicted in Fig. 7. Figure 7 shows that the inversion point increases with increasing the black hole mass. In Figs. 9 and 10 the plots of the inversion curve $P_i-T_i$ with different parameters are depicted.
\begin{figure}[h]
\includegraphics {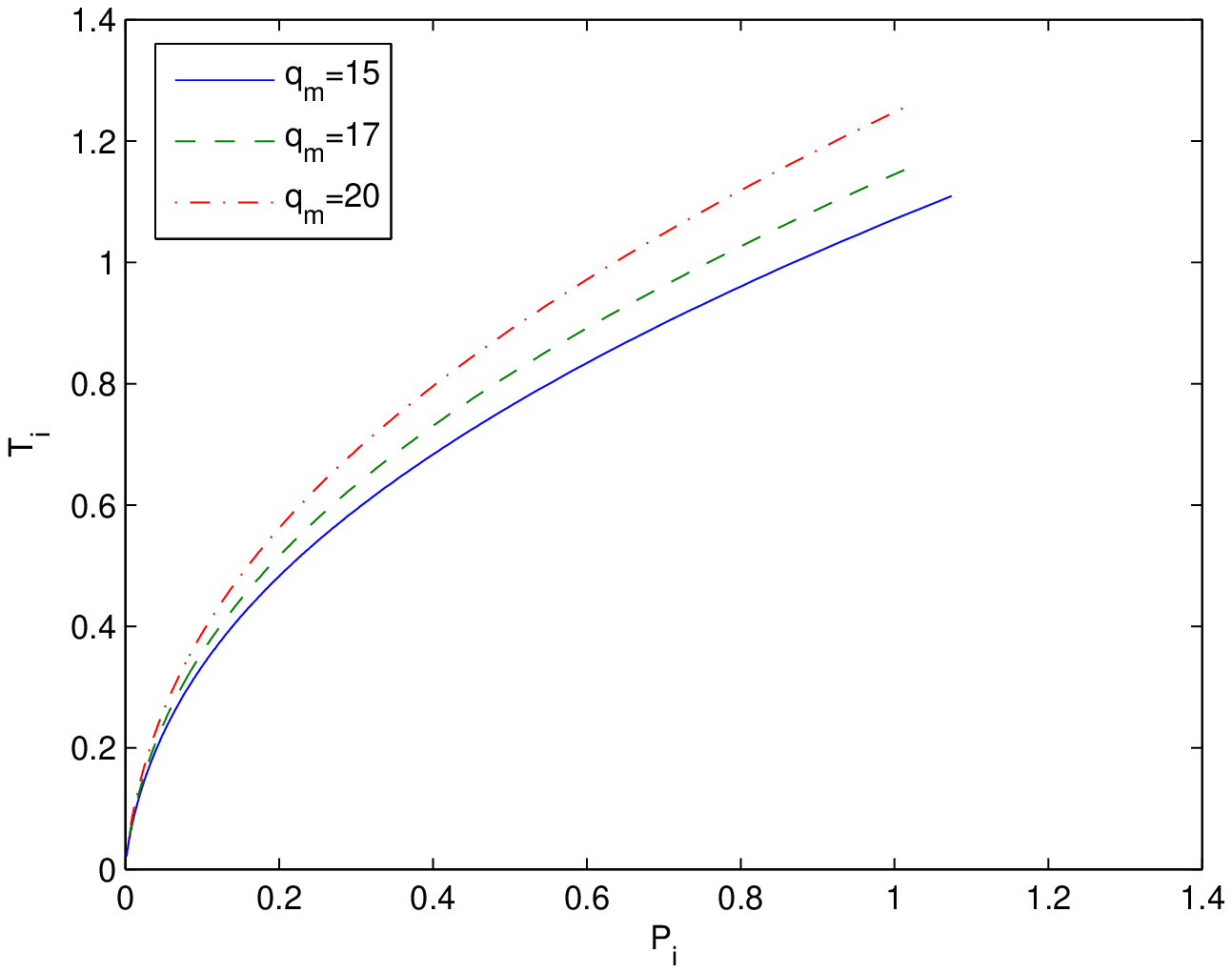}
\caption{\label{fig.9} The plots of the inversion temperature $T_i$ vs. pressure $P_i$ for $q_m=15$, $17$ and $20$, $\beta=1$. When magnetic charge $q_m$ increases the inversion temperature also increases.}
\end{figure}
\begin{figure}[h]
\includegraphics {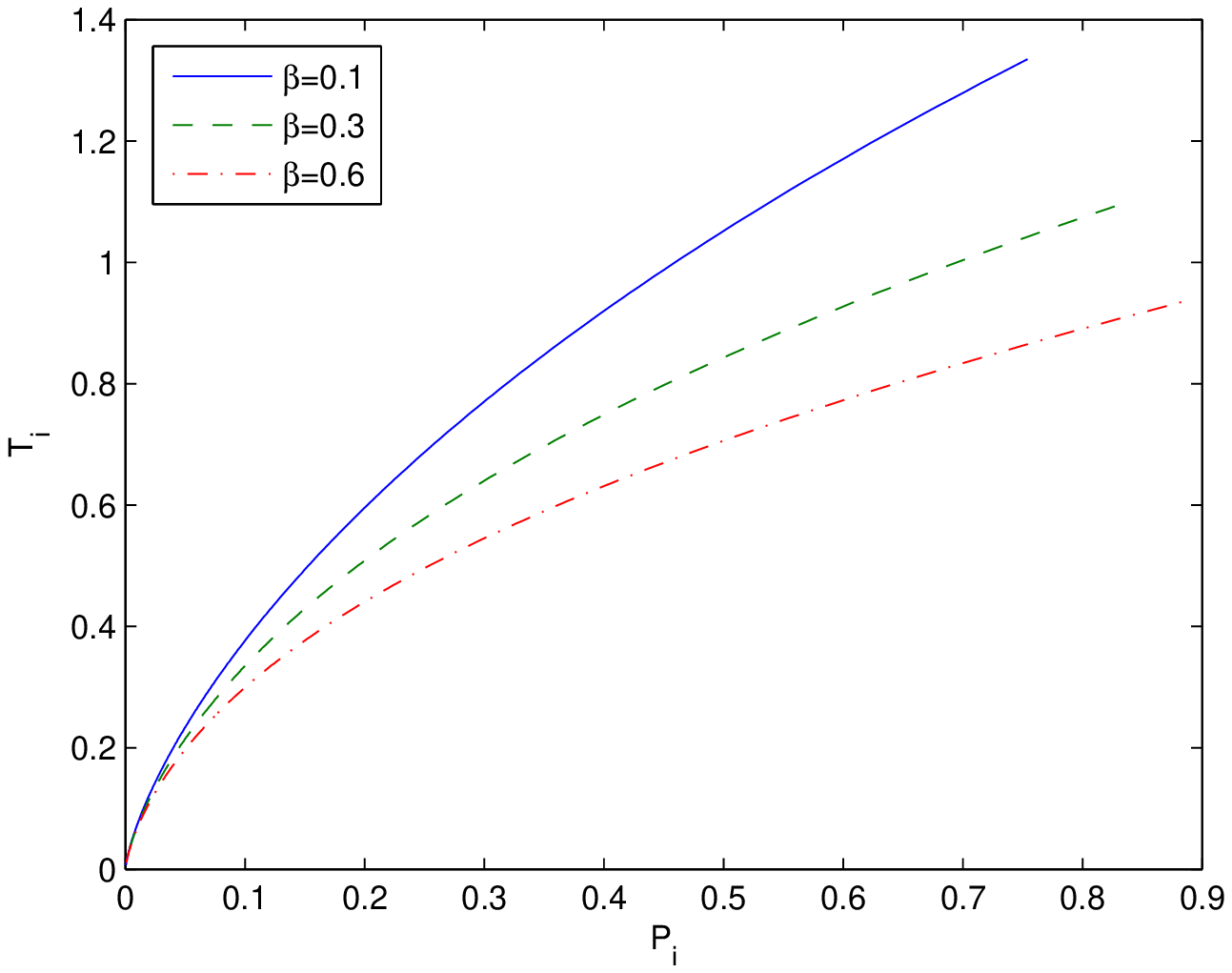}
\caption{\label{fig.10} The plots of the inversion temperature $T_i$ vs. pressure $P_i$ for $\beta=0.1$, $0.3$ and $0.6$, $q_m=10$. Figure shows that when the NED parameter $\beta$ increases then the inversion temperature decreases.}
\end{figure}
Figure 9 shows that with increasing magnetic charge $q_m$, for fixes coupling $\beta$ of black holes, the inversion temperature increases.
According to Fig. 10, when the coupling $\beta$ increases at fixed magnetic charge $q_m$, the inversion temperature decreases.
Making use of Eqs. (40), (41) and (38)  we obtain the Joule--Thomson coefficient
\[
\mu_J=\frac{4r_+\left(2\beta^{1/4}(r_+-6M)+A
+ B\right)}{3\left(4\beta^{1/4}(r_+-3M)+A
+C\right)},
\]
\[
A=3q_m^{3/2}\left[\pi-2\arctan\left(\frac{r_+}{\beta^{1/4}\sqrt{q_m}}\right)\right],
\]
\begin{equation}
B=\frac{2q_m^2\beta^{1/4}r_+ (3r_+^2+2q_m\sqrt{\beta})}{(r_+^2+q_m\sqrt{\beta})^2},~~~~C=\frac{2q_m^2\beta^{1/4}r_+}{r_+^2+q_m\sqrt{\beta}}.
\label{47}
\end{equation}
When the Joule--Thomson coefficient is positive, $\mu_J>0$, a cooling process occurs, while if $\mu_J<0$, a heating process holds. In Fig. 7 the area with $\mu_J>0$ is in the left side of inversion temperature borderline and $\mu_J<0$ corresponds to the right of borderline $T_i$.

\section{Conclusion and summary}

In this paper, we have obtained NED-AdS magnetic black hole solution and found metric and mass functions and their asymptotic. It was shown that black holes possess only one horizon. Corrections to the Reissner--Nordstr\"{o}m solution have been found which are in the order of ${\cal O}(r^{-4})$. We showed that when NED parameter $\beta$ increases at constant magnetic charge $q_m$, the event horizon radius $r_+$ decreases, however if magnetic charge increases at constant $\beta$, the event horizon radius increases.
The thermodynamics of NED-AdS black holes in an extended thermodynamic phase space has been studied. The cosmological constant plays the role of a thermodynamic pressure and the mass of the black hole is treated as the chemical enthalpy.  We found a thermodynamic quantity ${\cal B}$ conjugated to NED parameter $\beta$ and thermodynamic potential $\Phi_m$ conjugated to magnetic charge $q_m$. It was shown that the first law of black hole thermodynamics and the generalized Smarr formula hold. The analogy with the Van der Walls liquid–gas system with the specific volume was shown. We have calculated the Gibbs free energy and showed that first-order phase transitions occur and critical temperature and pressure were found. It was demonstrated that at the limit $\beta\rightarrow\infty$ we come to the Schwarzschild-AdS black holes.
The critical ratio $\rho_c=3/8+{\cal O}(\beta)$ have been obtained with the Van der Waals value critical ratio $3/8$.  The critical exponents equal to the Van der Waals system exponents. We have studied cooling and heating phase transitions of NED-AdS black holes through the Joule--Thomson adiabatic expansion. Isenthalpic $P-T$ diagrams and the inversion temperature curve were plotted. We found the dependence of the inversion temperature on magnetic charge and NED coupling of black holes depicted in figures 9 and 10. The inversion temperature curve separates the isenthalpic plots into two branches corresponding to cooling ($\mu_J>0$) and heating ($\mu_J<0$) processes of black holes at the Joule--Thomson adiabatic expansion.

\vspace{3mm}
\textbf{Appendix}
\vspace{3mm}

The Kretschmann scalar is given by
\begin{equation}
K(r)\equiv R_{\mu\nu\alpha\beta} R^{\mu\nu\alpha\beta}=
(f''(r))^2+\left(\frac{2f'(r)}{r}\right)^2+\left(\frac{2(f(r)-1)}{r^2}\right)^2.
\label{48}
\end{equation}
Making use of Eq. (12) at $G_N=1$ we obtain the derivatives
\[
f'(r)=\frac{2m_0}{r^2}-\frac{q_m^2}{r(r^2+q_m\sqrt{\beta})}+\frac{q_m^{3/2}}{\beta^{1/4}r^2}\arctan\left(\frac{r}{\beta^{1/4}\sqrt{q_m}}\right)
+\frac{2r}{l^2},
\]
\[
f''(r)=-\frac{4m_0}{r^3}+\frac{q_m^2}{r^2(r^2+q_m\sqrt{\beta})}-\frac{2q_m^{3/2}}{\beta^{1/4}r^3}\arctan\left(\frac{r}{\beta^{1/4}\sqrt{q_m}}\right)
\]
\begin{equation}
+\frac{q_m^2(3r^2+q_m\sqrt{\beta})}{r^2(r^2+q_m\sqrt{\beta})^2}+\frac{2}{l^2}.
\label{49}
\end{equation}
With the help of Eqs. (12) and (49) we plotted  the Kretschmann scalar versus $r$.
The Kretschmann scalar as $r\rightarrow 0$ goes to infinity, and therefore, there is a space-time singularity at $r=0$. As $r\rightarrow \infty$ the Kretschmann scalar approaches to constant due to cosmological constant.
\begin{figure}[h]
\includegraphics{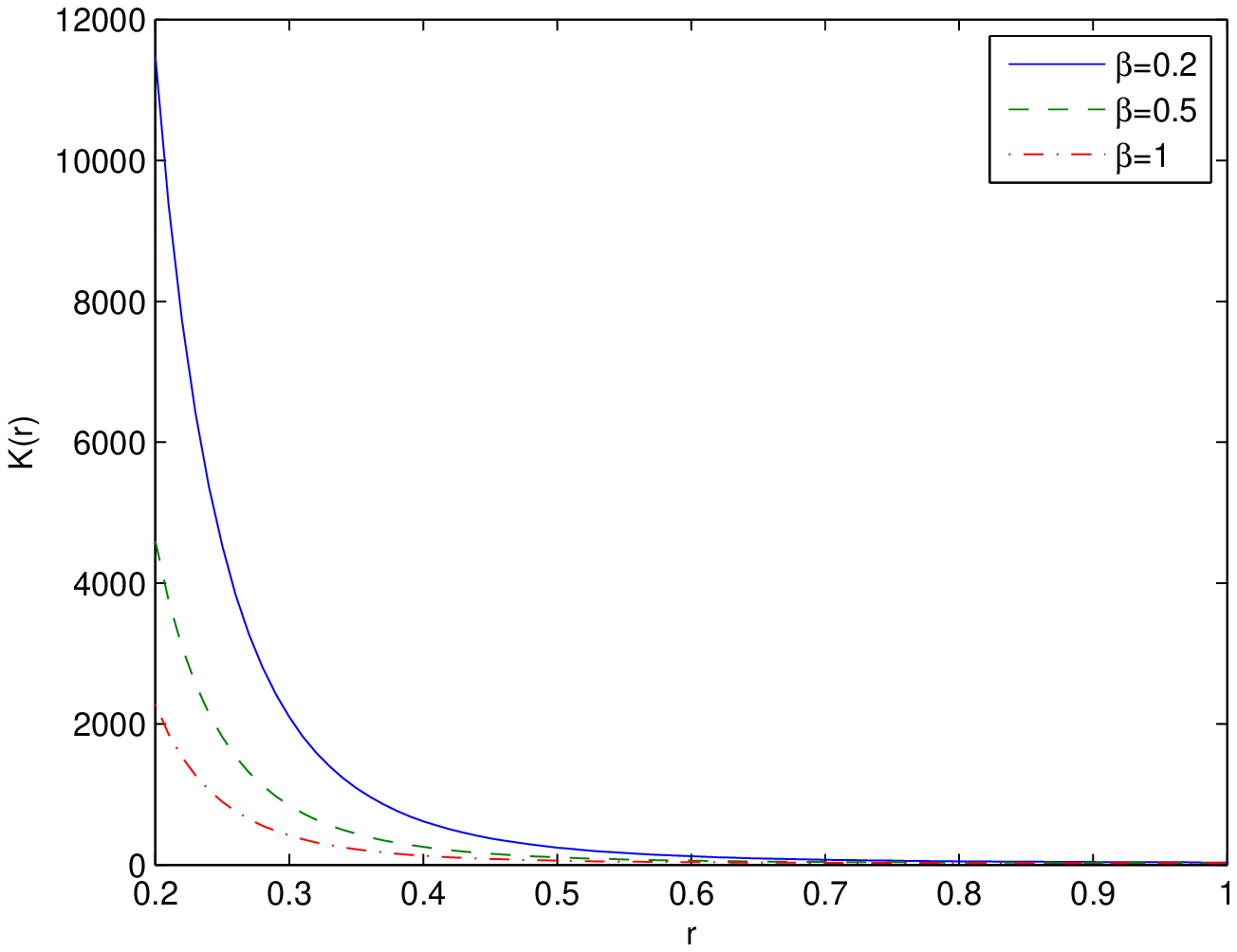}
\caption{\label{fig.11}The plot of the function $K(r)$ vs. $\beta$ for $l=1$, $m_0=0$. The solid line corresponds to $\beta=0.2$, the dashed line corresponds to $\beta=0.5$ and the dashed-dotted line corresponds to $\beta=1$. As $r\rightarrow 0$ the Kretschmann scalar goes to infinity. Thus, there is a space-time singularity at $r=0$. As $r\rightarrow \infty$ the Kretschmann scalar approaches to constant due to cosmological constant.}
\end{figure}
Figure 11 shows that the curvature invariant $K(r)$ is not bounded even for $m_0=0$.

\end{document}